\documentclass[fleqn,usenatbib]{mnras}

\usepackage{newtxmath}

\usepackage[T1]{fontenc}
\usepackage{ae,aecompl}

\usepackage{graphicx}	
\usepackage{amsmath}	
\usepackage{amssymb}	

\usepackage[dvipsnames]{xcolor}
\hypersetup{
  colorlinks   = true, 
  urlcolor     = RoyalBlue, 
  linkcolor    = RoyalBlue, 
  citecolor   = MidnightBlue 
}
\usepackage{natbib}
\usepackage{color}
\usepackage{rotating}
\usepackage{url}
\usepackage{ifthen}

\usepackage{bm}
\usepackage{bbold}
\usepackage{aas_macros}
\usepackage{verbatim}
\usepackage{algorithm}
\usepackage{algpseudocode}
\usepackage{enumitem}
\usepackage{subfig}

\DeclareMathAlphabet{\mathpzc}{OT1}{pzc}{m}{it}

\newcommand{\mvec}[1]{\bm{#1}}

\newcommand{\myvec}[1]{\mvec{#1}}

\newcommand\numberthis{\addtocounter{equation}{1}\tag{\theequation}}

\SetSymbolFont{symbols}{bold}{OMS}{cmsy}{b}{n}
\DeclareSymbolFont{bmisymbols}{OML}{cmm}{b}{it}

\title[Super-resolution emulated simulations]{Super-resolution emulator of cosmological simulations using deep physical models}

\author[D. Kodi Ramanah et al.]{Doogesh Kodi Ramanah,$^{1,2,3}$\thanks{ramanah@nbi.ku.dk} Tom Charnock,$^{2}$\thanks{charnock@iap.fr} Francisco Villaescusa-Navarro,$^{4,5}$\newauthor Benjamin D. Wandelt$^{2,3,5}$\\
$^{1}$ DARK, Niels Bohr Institute, University of Copenhagen, Lyngbyvej 2, 2100 Copenhagen, Denmark\\
$^{2}$ Sorbonne Universit\'e, CNRS, UMR 7095, Institut d'Astrophysique de Paris, 98 bis bd Arago, 75014 Paris, France\\
$^{3}$ Sorbonne Universit\'e, Institut  Lagrange  de  Paris  (ILP),  98  bis bd Arago, 75014 Paris, France\\
$^{4}$ Department of Astrophysical Sciences, Princeton University, Peyton Hall, Princeton NJ 08544-0010, USA\\
$^{5}$ Center for Computational Astrophysics, Flatiron Institute, 162 5th Avenue, 10010, New York, NY, USA\\
}

\date{Accepted XXX. Received YYY; in original form ZZZ}

\pubyear{2020}

\begin{document}
\label{firstpage}
\pagerange{\pageref{firstpage}--\pageref{lastpage}}
\maketitle

\begin{abstract}
We present an extension of our recently developed Wasserstein optimized model to  emulate accurate high-resolution features from computationally cheaper low-resolution cosmological simulations. Our deep physical modelling technique relies on restricted neural networks to perform a mapping of the distribution of the low-resolution cosmic density field to the space of the high-resolution small-scale structures. We constrain our network using a single triplet of high-resolution initial conditions and the corresponding low- and high-resolution evolved dark matter simulations from the \textsc{Quijote} suite of simulations. We exploit the information content of the high-resolution initial conditions as a well constructed prior distribution from which the network emulates the small-scale structures. Once fitted, our physical model yields emulated high-resolution simulations at low computational cost, while also providing some insights about how the large-scale modes affect the small-scale structure in real space.
\end{abstract}




\section{Introduction}
\label{intro}

One of the most powerful ways to learn about the cosmological model of our Universe relies on extracting the physical information encoded in its large-scale structure. However, the highly non-linear dynamics involved in gravitational structure formation renders this a complex problem, such that numerical simulations are a prerequisite to obtain theoretical predictions in the fully non-linear regime of structure growth. Cosmological inference from next-generation galaxy surveys such as Euclid \citep{euclid2011report} and the Large Synoptic Survey Telescope (LSST) \citep{lsst2008summary} will require a large number of detailed and high-resolution simulations to generate mock observations of the Universe. Fast and reliable emulators of these complex dynamics would, therefore, be crucial for data analysis and light cone production for such survey missions.

\medskip
By virtue of their versatility and effectiveness, deep generative models have recently been developed for a range of applications in cosmology, especially with recent advances in the field of deep learning. \citet{he2019learning} constructed a deep neural network, based on the U-Net \citep{ronneberger2015UNet} architecture, to predict the non-linear cosmic structure formation from linear perturbation theory. \citet{giusarma2019learning} employed a similar architecture to map standard $N$-body simulations to non-standard ones with massive neutrinos. \citet{zhang2019darkmatter} developed a two-phase convolutional neural network to predict the galaxy distribution in hydrodynamic simulations from their corresponding 3D dark matter density fields. Various approaches based on 3D convolutional neural networks have been developed to generate mock halo catalogues from the cosmological initial conditions \citep{berger2018volumetric, bernardini2019predicting}. The deep convolutional generative adversarial network architecture \citep{goodfellow2014generative, radford2015unsupervised} and its variants \citep{arjovsky2017wgan} have been used to generate cosmological weak lensing convergence maps with high statistical confidence \citep{mustafa2017creating, tamosiunas2020toward}, realistic 2D and 3D realizations of the cosmic web \citep{rodriguez2018fast, perraudin2019cosmological, feder2020nonlinear}, 3D gas pressure distributions from dark matter fields \citep{troster2019painting}, and 3D cosmic neutral hydrogen (HI) distributions \citep{zamudio2019HIgan}. Recent deep U-Net models were also successful in generating 3D gas density distributions with dark matter annihilation feedback \citep{list2019black}.

\medskip
In this work, we present an extension of our recent Wasserstein optimized network, developed for mapping approximately evolved dark matter fields to their corresponding 3D halo count distributions obtained from $N$-body simulations \citep{DKR2019painting}. Here, we accurately emulate super-resolution $N$-body simulations using their respective initial conditions and evolved low-resolution structures. We find, as with the previous work, that using a well motivated prior and simple physical principles results in high computational efficiency and remarkable performance while requiring only one set of simulations for fitting. 

\medskip
The remainder of this paper is structured as follows. In Section~\ref{quijote_sims}, we outline the salient features of the \textsc{Quijote} $N$-body simulation suite employed in the fitting and validation of our neural network which is described in Section~\ref{deep_generative_model}. We briefly review the conceptual foundations of the Wasserstein generative networks in Section~\ref{WGN}, followed by a description of the network architecture and optimization procedure in Sections~\ref{neural_network_architecture} and \ref{fitting_methodology}, respectively. We subsequently evaluate the performance of our high-resolution $N$-body emulator in terms of its capacity to reproduce the complex structures of the cosmic web from a ground truth simulation using various metrics in Section~\ref{results}. Finally, in Section~\ref{conclusions}, we provide a summary of the key aspects of our work, possible avenues for future investigations and potential applications where employing such an emulator may yield crucial advantages.

\section{\textsc{Quijote} simulations}
\label{quijote_sims}

The \textsc{Quijote} suite of simulations \citep{paco2019quijote} is a set of 43100 full $N$-body simulations encompassing over 7000 different cosmological models, specifically designed to provide high quality data to train machine learning algorithms. Here, we use only one high-resolution simulation, one low-resolution simulation generated with the same initial conditions and those initial conditions as our data for fitting the model. We use a larger set of these triplets from the remainder of the suite to validate and test the model.

\medskip
The simulations cover a cosmological volume of 1000$h^{-1}$~Mpc and are initialized at a redshift $z=127$ and evolved to present time using the TreePM code \textsc{gadget3} -- an improved version of \textsc{gadget2} \citep{springel2005gadget2}. The gravitational softening length is set to $1/40$ of the mean interparticle separation, corresponding to 100$h^{-1}$~kpc and 50$h^{-1}$~kpc, for the low- and high-resolution simulations with $256^3$ and $512^3$ particles tracing dark matter, respectively. The input matter power spectrum and transfer functions are obtained by rescaling their present-day ($z=0$) counterparts from \textsc{camb} \citep{camb1999}. The simulations assume a Planck-like $\Lambda$CDM cosmology \citep{planck2018cosmo} with $\Omega_\text{m}=0.3175$, $\Omega_\text{b}=0.049$, $h=0.6711$, $\sigma_8=0.834$, $n_{\mathrm{s}}=0.9624$ and $w=-1$. A comprehensive description of the \textsc{Quijote} simulations is provided in \citet{paco2019quijote}.

\section{Deep generative modelling}
\label{deep_generative_model}

\begin{figure*}
	\centering
		{\includegraphics[scale=0.625,clip=true]{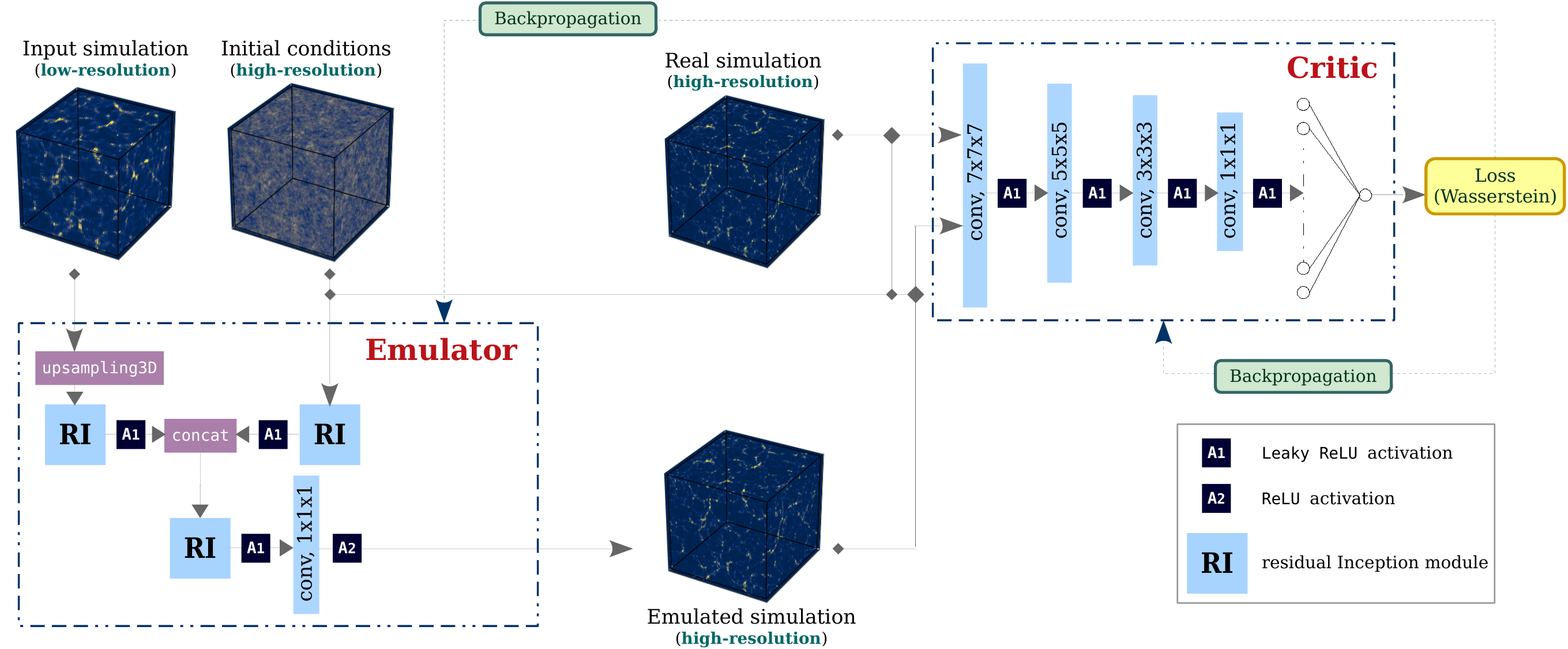}}
	\caption{Schematic representation of the super-resolution emulator implemented in this work. The emulator approximates the underlying mapping of the distribution of low-resolution density field to high-resolution structures, with the input initial conditions providing an informative prior distribution from which the emulator constructs the fine structures, to yield a super-resolution field. The difference between the output of the critic for the real and emulated density fields, conditional on the initial conditions, is the approximate Wasserstein distance, which is minimized to fit the super-resolution $N$-body emulator.}
	\label{fig:WGN_schematic}
\end{figure*}

\subsection{Wasserstein generative network}
\label{WGN}

The Wasserstein generative model is an improved variant of generative adversarial networks \citep[GAN,][]{goodfellow2014generative}. GANs cast the generative process as a competition between a generator $\mathcal{G}_{\theta}$ and a discriminator $\mathcal{D}_{\varphi}$, where $\theta$ and $\varphi$ correspond to the weights of the generator and discriminator, respectively. The role of the former is to produce some artificial data given a vector of random noise, while the latter must differentiate between the generated samples and the \emph{real} ones. The Wasserstein variant \citep{arjovsky2017towards} works similarly with a slight change in the loss function, but with a large change in interpretation of \emph{training}. Instead of training a discriminative classifier, one approximates the optimal transport distance which takes the samples from the generated distribution $\myvec{g}=\mathcal{G}_\theta(\myvec{z})$ with $\myvec{z}\sim\mathbb{P}_\mathrm{z}$ to the target distribution $\mathbb{P}_\mathrm{r}$ using a critic network $\mathcal{C}$. This optimal transport distance is known as the Wasserstein-1 distance or Earth mover's distance. A more in-depth description of the conceptual framework underlying the Wasserstein generative network is provided in our previous work \citep[][hereafter KCL19]{DKR2019painting}.

\medskip
Computing the Wasserstein-1 distance, in practice, is intractable, but an approximation is feasible for the case where the critic network $\mathcal{C}_\varphi$ is a 1-Lipschitz function parameterized by weights $\varphi$ which lie in a compact space. To enforce this constraint, one can introduce a gradient penalty to the loss function, as proposed by \citet{gulrajani2017improved}, which provides a substantial improvement over previous weight-clipping methods. The augmented loss function for the Wasserstein generative model may be expressed as
\begin{align*}
 \mathcal{L} =  \underset{\myvec{z} \sim \mathbb{P}_\mathrm{z}}{\mathbb{E}} \left[ \mathcal{C}_\varphi(\mathcal{G}_\theta(\myvec{z})) \right] &- \underset{\myvec{x} \sim \mathbb{P}_{\mathrm{r}}}{\mathbb{E}} \left[ \mathcal{C}_\varphi (\myvec{x}) \right]  \\ &+ \lambda \underset{\hat{\myvec{x}} \sim \mathbb{P}_{\hat{\myvec{x}}}}{\mathbb{E}} \left[ \left( || \nabla_{\hat{\myvec{x}}} \mathcal{C}_\varphi (\hat{\myvec{x}}) ||_2 - 1 \right)^2 \right] , \numberthis
\label{eq:augmented_loss_gradient_penalty_GAN}
\end{align*}
where the vector $\myvec{z} \sim \mathbb{P}_{\mathrm{z}}$ is a sample of the set of low-resolution evolved dark matter density field and corresponding high-resolution initial conditions which emulates a high-resolution evolved dark matter density simulation $\myvec{g}=\mathcal{G}_\theta(\myvec{z})$. $\myvec{x}\sim\mathbb{P}_\mathrm{r}$ is a sample from the true distribution of high-resolution evolved dark matter density fields. The first two terms yield the approximate Wasserstein distance while the last term ensures that the gradient of the critic network remains close to unity. $\hat{\myvec{x}} \sim \mathbb{P}_{\hat{\myvec{x}}}$ is obtained by interpolating between the real and generated samples: $\hat{\myvec{x}} = \varepsilon \myvec{x} + (1 - \varepsilon)\myvec{g}$ for $\varepsilon$ sampled randomly and uniformly, $\varepsilon \in [0,1]$, and $\lambda$ is an arbitrary penalty coefficient.

\subsection{Neural network architecture}
\label{neural_network_architecture}

A schematic representation of our super-resolution emulator and the critic is illustrated in Fig.~\ref{fig:WGN_schematic}. As with our halo painting network in KLC19, we consider the \emph{generator} as an \emph{emulator}, i.e. the input to the network is not a flat array of noise, as usually considered for generative models, but rather involves 3D boxes of low-resolution density field and corresponding high-resolution initial conditions. As such, the super-resolution emulator performs a physical mapping from the distribution of low-resolution density field to high-resolution structures, conditional on the initial conditions. The critic compares the difference between a scalar summary of the features present in the real and emulated high-resolution fields. By minimizing the difference between these scalar summaries, we fit the function which transports the set of features in the emulated field to the real simulation. This is the approximate Wasserstein distance. Once the parameters of this critic have converged, it can be used as the metric (loss function) which is minimized to fit our super-resolution $N$-body emulator.

\medskip
The choice of architecture for our high-resolution emulator and the critic is depicted in Fig.~\ref{fig:WGN_schematic}. We employ residual Inception blocks (cf. Fig.~4 in KLC19 for a schematic illustration), as outlined below, in designing the emulator, with the additional inclusion of 3D convolutions with kernel size of $7\times7\times7$. All the convolutional layers, including those in the Inception blocks, have six filter channels, except for the output layer which has a single filter {\ttfamily ReLU} activation. The {\ttfamily leaky ReLU} activation function with a leaky parameter of $\alpha=0.1$ encodes the required approximate non-linearity in the initial layers. We used the same critic architecture as in KLC19, except that the initial conditions are concatenated with the input real high-resolution simulation. The critic contains four 3D convolutional layers with {\ttfamily leaky ReLU} activation, with an initial kernel of $7\times7\times7$, and gradually reducing the kernel size to $1\times1\times1$, with the subsequent output flattened and fed to a fully connected layer with linear activation.

\medskip
The residual Inception \citep{szegedy2017residualinception} module combines the Inception \citep{szegedy2015going, szegedy2016rethinking} architecture, which performs parallel convolutions with kernels of distinct sizes, and residual connections \citep{he2016resnet} between the input to the module to its subsequent output, thereby facilitating the flow of small-scale information. This residual Inception block allows feature extraction on different scales of the density field simultaneously and significantly improves computational efficiency.

\medskip
The super-resolution emulator is designed to perform a physical mapping that encodes some fundamental symmetries as motivated by basic physical principles and the cosmological principle. In particular, we enforce translational invariance via the use of 3D convolutions and the random flip of the inputs during training, and rotational invariance by performing a random rotation of the inputs while optimizing the network. The (non-) locality of the super-resolving procedure is inbuilt in the largest convolutional kernel in the Inception module, whose size is motivated by causal transport arguments outlined in the next paragraph. Finally, the non-linear activations provide the adequate non-linearity as required by this physical mapping.

\medskip
The typical displacement field, derived from the initial conditions, is $\sim$ 5$h^{-1}$~Mpc in a $\Lambda$CDM Universe with Planck cosmological parameters \citep{planck2018cosmo}, with an upper limit of $\sim$ 20$h^{-1}$~Mpc for the fastest moving objects, as substantiated in \citet{lavaux2019systematic}. The $7\times7\times7$ kernel in the first residual Inception block has a receptive field of $\sim$ 24$h^{-1}$~Mpc and ensures that the network is capable of encapsulating the causal displacements between the high-resolution initial conditions and evolved density field. This is necessary to transport the initial conditions to the correct regions of the emulated field. This transportation is provided by the second Inception module, after concatenating the subsequent feature maps. The effective receptive patch of nearly 50$h^{-1}$~Mpc is sufficiently large to pull all the relevant information from the respective inputs. The largest scale modes of the emulated density field are provided by the evolved low-resolution simulation. Although the network is never fit with the large modes explicitly (with extremely large convolutional kernels), they are incorporated by fitting with small sub-patches of the low-resolution field which includes these modes on average. In Section~\ref{filter_visualization}, we examine a selection of kernels and feature maps in an attempt to introspect our super-resolution simulation emulator.

\subsection{Fitting methodology}
\label{fitting_methodology}

We make use of a single set of three 3D gridded dark matter fields for inferring the parameters of the emulator, corresponding to the high-resolution initial conditions, and low- and high-resolution density fields evolved from these initial conditions (at $z=0$). The low- and high-resolution simulation boxes have $256^3$ and $512^3$ voxels, respectively, and they both have a size of 1000$h^{-1}$~Mpc. We utilize most of the boxes for fitting, with a independent portion kept for validation. The optimization rationale entails minimizing the approximate Wasserstein distance between the true and emulated high-resolution density fields, conditional on the initial conditions, such that the super-resolution simulation emulator approximates the correct mapping from low- to high-resolution density fields.

\medskip
The optimization routine proceeds as follows: a set of two randomly chosen 3D patches of high-resolution initial conditions and the corresponding patch of low-resolution density field, with respective sizes of $40^3$ and $20^3$ voxels, are passed through the first network to emulate a high-resolution density volume, with the input patches encoding a sufficiently large number of informative features for optimizing the network. The physical voxel sizes for the low- and high-resolution fields are $\sim$ 4$h^{-1}$~Mpc and $\sim$ 2$h^{-1}$~Mpc, respectively. As such, the sub-volume involved during the optimization procedure is $\sim$ 80$h^{-1}$~Mpc. The convolutional layers in the residual Inception module do not utilize any padding so as not to induce any numerical artefacts arising from the boundaries of the input 3D slices. Hence, in order to eliminate the need for padding, the input size is conveniently chosen to be larger, such that the network output has the desired box size. For our particular network architecture, as depicted in Fig.~\ref{fig:WGN_schematic}, since there are two Inception modules, with the largest convolutional kernel being $7\times7\times7$, the input slices must be larger by $(7-1) \times 2 = 12$ voxels on each side. During optimization, the emulator, therefore, physically maps a sub-volume of $\sim$ 80$h^{-1}$~Mpc to $\sim$ 55$h^{-1}$~Mpc. Once the model is optimized, the inputs may be of any arbitrary box size, exploiting the translational invariance of convolutional kernels, to predict the high-resolution density field of corresponding desired size.

\medskip
To artificially augment the data manifold, the 3D volumes are randomly rotated and mirrored along the three axes on entry to the emulator. The emulated super-resolution field and the corresponding true high-resolution density volume (rotated and mirrored in the same way as the initial conditions and the low-resolution simulation) are then passed through the critic. The parameters of the critic are optimized to minimize the difference between the summary of the true density field and the emulated one via equation~\eqref{eq:augmented_loss_gradient_penalty_GAN}, which equates to finding the optimal transport of features conditional on the architecture of the critic. At this step, the parameters of the emulator are kept fixed so that the critic properly approximates the difference between the features of the emulated samples $\myvec{g}=\mathcal{G}_\theta(\myvec{z})$ from distribution $\mathbb{P}_\mathrm{z}$ and true distribution $\mathbb{P}_\mathrm{r}$ via the optimization through several steps until convergence. Once convergence of the critic is reached, a single weight update of the emulator is made using 
\begin{equation}
    \mathcal{L}_\textrm{emulator}=\underset{\myvec{z} \sim \mathbb{P}_\mathrm{z}}{\mathbb{E}}[\mathcal{C}_\varphi(\mathcal{G}_\theta(\myvec{z}))].
\end{equation}
The batch size is initialized at unity and is doubled for every subsequent one hundred thousand weight updates to make better estimates of the expectation value of the distribution once each individual emulated simulation is close to its corresponding true simulation \citep{smith2017batch}. This iterative optimization routine is repeated until convergence of the two distributions.

\medskip
The neural network and optimization procedure are implemented in \textsc{TensorFlow} \citep{abadi2016tensorflow}. We set the number of iterations for the critic updates to $n_{\mathrm{critic}} = 10$ and the coefficient for the gradient penalty to $\lambda = 10$. We employ the {\it Adam} \citep{kingma2014adam} optimizer, with a learning rate of $\eta=10^{-4}$ and first and second moment exponential decay rates of $\beta_1=0.5$ and $\beta_2=0.999$, respectively. We optimize the emulator for $\sim\,5\times10^5$ weight updates, requiring around 120 hours on a NVIDIA Quadro P6000. As with the halo painting network from our previous work (cf. KLC19), building a physically motivated neural engine, in contrast to conventional extremely deep black box approaches, yields crucial advantages in terms of relatively low network complexity. As a consequence, much less data is needed to fit the emulator. Our high-resolution emulator has $\mathcal{O}(10^4)$ network parameters, and is fit with a data set of only one each of the initial conditions, low- and high-resolution density fields. This is extremely cost efficient in comparison to recent deep generative models developed in cosmology with $\mathcal{O}(10^7)$ parameters, which require several thousands of numerical simulations for training \citep[e.g.][]{rodriguez2018fast, he2019learning, giusarma2019learning}. Physically motivated neural networks (cf. KLC19 and this work) deliver results of at least comparable accuracy, as we will show in the following.

\section{Results}
\label{results}

\begin{figure}
	\centering
		{\includegraphics[width=\hsize,clip=true]{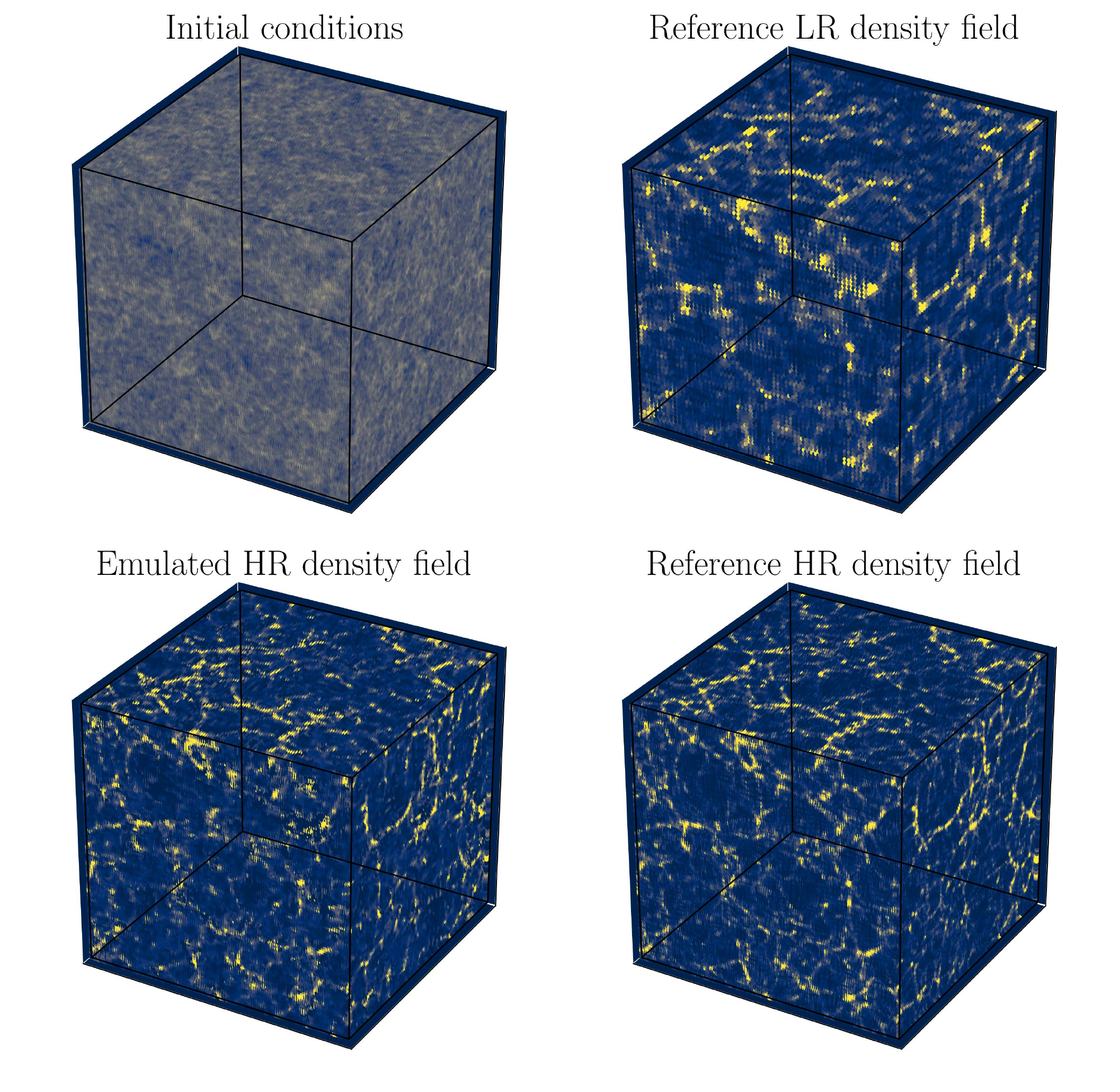}}
	\caption{High-resolution (HR) prediction ({\it bottom-left panel}) of the non-linearly evolved density field by our super-resolution emulator for a 3D slice of length 500$h^{-1}$~Mpc. For comparison, the reference density field from the high-resolution simulation is depicted in the bottom-right panel, with the corresponding input initial conditions and low-resolution (LR) density field illustrated in the top panels. This provides a qualitative assessment of the efficacy of the high-resolution emulator.}
	\label{fig:visual_inspection}
\end{figure}

We assess the performance of the super-resolution emulator using a series of both qualitative and quantitative diagnostics. Fig.~\ref{fig:visual_inspection} depicts the 3D visualization of the model prediction and the ground truth for a 3D slice of length 500$h^{-1}$~Mpc. The corresponding input initial conditions and low-resolution density field are also provided for completeness. Visually, there is extremely good agreement between the network prediction and reference high-resolution simulation.

\medskip
For a more quantitative evaluation, we employ four summary statistics, namely the 1D probability distribution function (PDF), power spectrum, bispectrum and the void size function, to compare the properties of the emulated high-resolution 3D density field to their corresponding (reference) numerical simulations. Our test set consists of ten distinct (unseen) simulations and the following plots indicate the means and respective $1\sigma$ confidence regions. Finally, in an attempt to introspect the inner-workings of the emulator, we visualize the filters of the convolutional kernels and the resulting feature maps and comment on their meaning.

\subsection{1D probability distribution function}
\label{1D_PDF}

\begin{figure}
	\centering
		{\includegraphics[width=\hsize,clip=true]{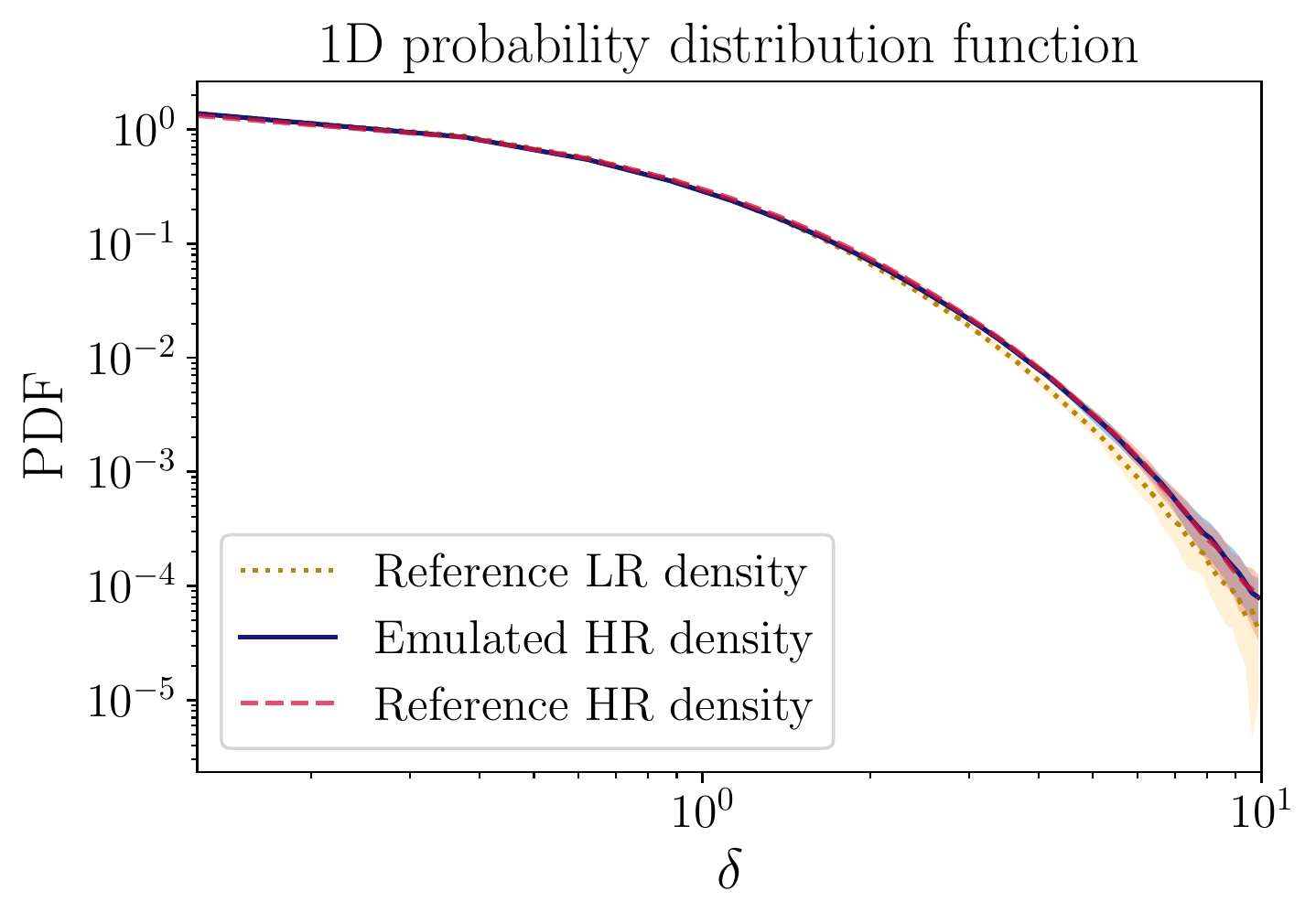}}
	\caption{Normalized probability distribution function (PDF) of the dark matter density contrast. The solid blue line indicates the mean for the ten emulated high-resolution (HR) simulations, which matches very closely that of the reference simulations depicted by the dashed red line. The shaded regions indicate their respective $1\sigma$ regions. For completeness, the 1D PDF of the corresponding low-resolution (LR) simulations is also shown.}
	\label{fig:PDF_1D}
\end{figure}

We verify whether the emulated high-resolution simulations manage to reproduce the 1D PDF of the real simulations. The 1D PDF describes the dark matter distribution across the voxels of the grid. We first apply a smoothing to the density field with a top-hat filter on a scale of 10$h^{-1}$~Mpc and compute the 1D PDF by binning the dark matter density contrast. The resulting distributions for the emulated and reference low- and high-resolution density fields are illustrated in Fig.~\ref{fig:PDF_1D}. As can be seen, there is an almost identical agreement between the emulated simulations and the ground truth.

\subsection{Power spectrum}
\label{power_spectrum}

\begin{figure}
	\centering
		{\includegraphics[width=\hsize,clip=true]{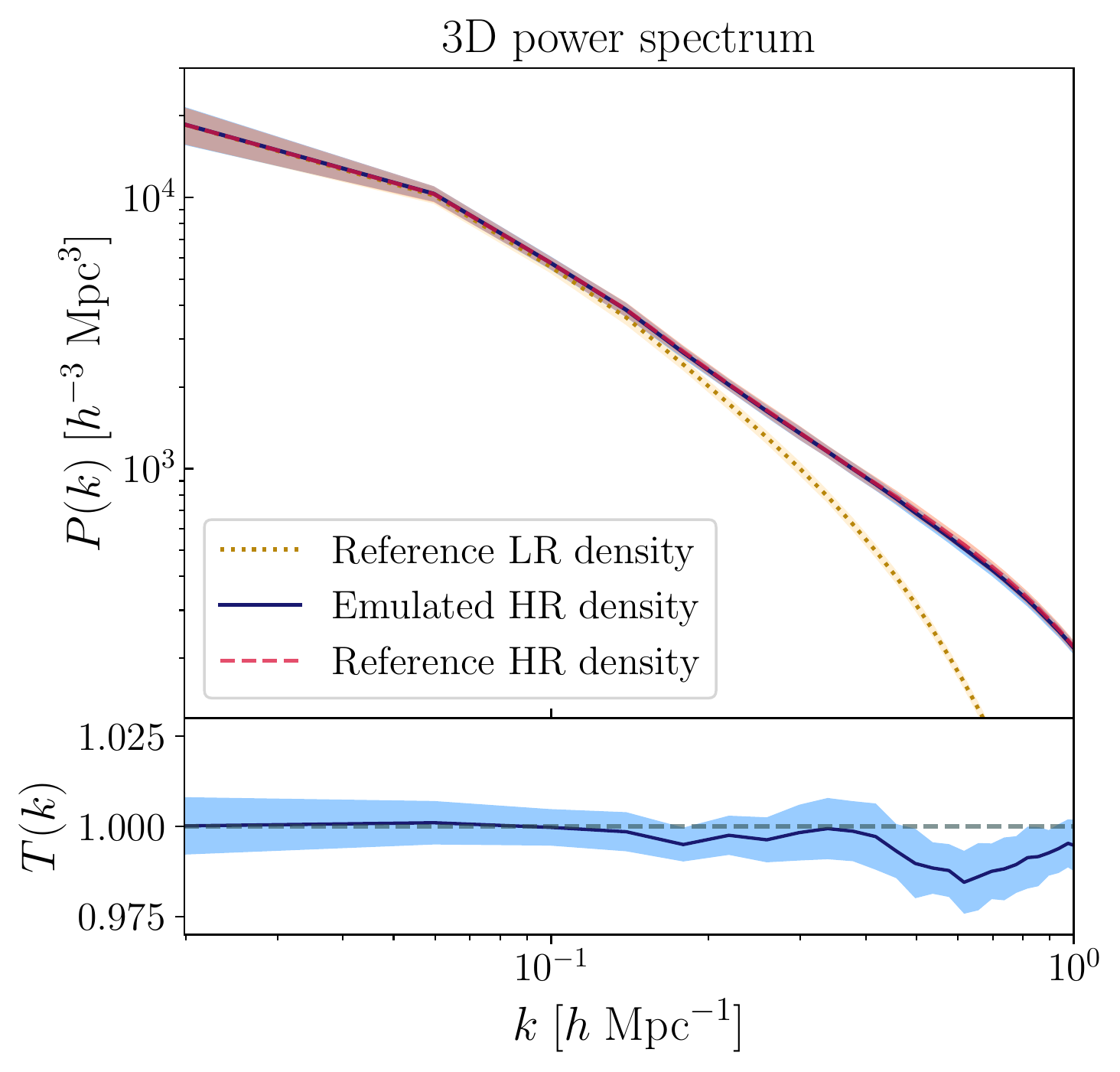}}
	\caption{Summary statistics of the 3D power spectra for the emulated high-resolution (HR) simulations and corresponding reference high- and low-resolution (LR) realizations. The solid lines indicate the mean for ten emulated HR realizations, with the shaded areas depicting the corresponding 1$\sigma$ confidence regions. {\it Top panel:} The power spectrum diagnostics demonstrate that the emulated HR density fields reproduce extremely well the characteristic two-point statistics of the reference HR fields. {\it Bottom panel:} The transfer function better quantifies the deviation, as a function of Fourier modes, of the emulated power spectra to the ground truth. This further illustrates the fidelity of the HR simulations from our super-resolution emulator at the level of two-point summary statistics.}
	\label{fig:Pk_Tk_subplot}
\end{figure}

Summary statistics are widely used in cosmology to compare model predictions with actual observations or to extract information from the latter. We first consider the two-point correlation function, $\xi(r)$, as a reliable metric to evaluate the capacity of our super-resolution emulator to encode the essential Gaussian information of the true high-resolution density field. Its Fourier transform, the power spectrum, denoted by $P(k)$, is defined as follows:
\begin{align}
    \xi(|\myvec{r}|) &= \langle \delta(\myvec{r}') \delta(\myvec{r}' + \myvec{r}) \rangle \label{eq:2PCF} \\
    P(|\myvec{k}|) &= \int \mathrm{d}^3 \myvec{r} \; \xi(\myvec{r}) e^{i\myvec{k}\cdot\myvec{r}} , \label{eq:power_spectrum}
\end{align}
where $k$ is the wavenumber of the fluctuation, i.e. $k = 2\pi / \lambda$, for given wavelength $\lambda$. The matter density distribution is typically described as a dimensionless over-density field (or density field contrast) via $\delta(\myvec{r}) = \rho(\myvec{r})/\bar{\rho}-1$, where $\rho(\myvec{r})$ is the matter density at position $\myvec{r}$ and $\bar{\rho}$ is the mean density. The above two-point statistics provide a sufficient statistical description of Gaussian random fields. Since the cosmic density field on large scales, or in the initial stages of structure growth, resemble a Gaussian field, such metrics are predominantly employed in standard cosmological analyses. The transfer function, $T(k)$, defined as the square root of the ratio of the emulated power spectrum, $P_\mathrm{g}(k)$, to the reference power spectrum, $P_{\rm{ref}}(k)$, 
\begin{equation}
    T(k) \equiv \sqrt[]{\frac{P_\mathrm{g}(k)}{P_{\rm{ref}}(k)}} ,
	\label{eq:transfer_function}
\end{equation}
more adequately characterizes the agreement between the amplitudes, as a function of Fourier modes.

\medskip
Fig.~\ref{fig:Pk_Tk_subplot} illustrates the power spectrum and transfer function for the emulated high-resolution simulations and corresponding reference simulations. In the top panel, the amplitude and shape of the emulated power spectra match the ground truth extremely well implying that the super-resolution emulated density field has the correct statistical properties at the level of two-point statistics. The transfer function, depicted in the bottom panel, is close to unity up to a scale of $k \sim 1.0$ $h^{-1}$~Mpc, with less than $1.5\%$ deviation, further substantiating the accuracy of the emulation even in the non-linear regime of cosmic structure formation.

\subsection{Bispectrum}
\label{bispectrum}

\begin{figure}
	\centering
		{\includegraphics[width=\hsize,clip=true]{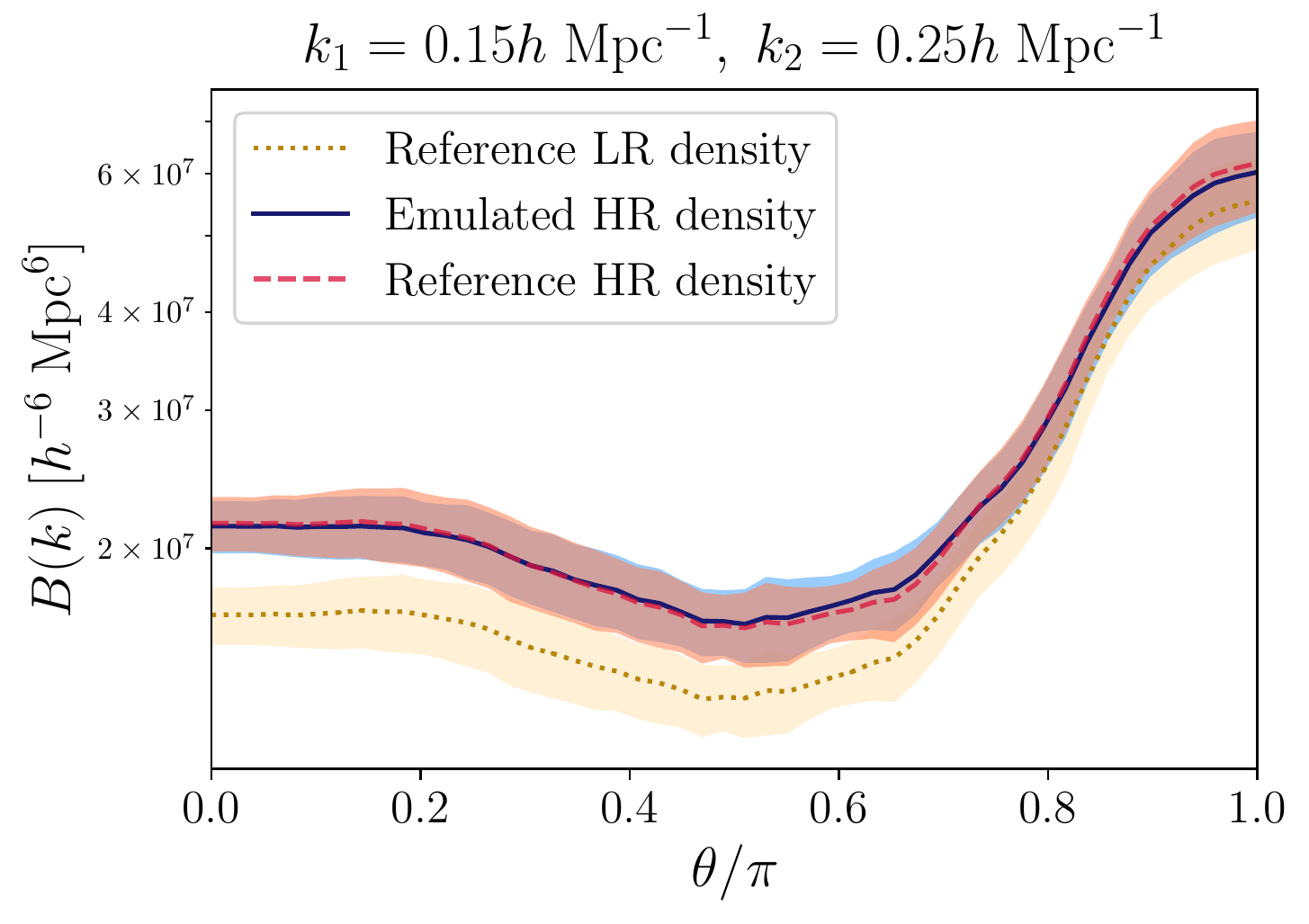}}
	\caption{Summary statistics of the 3D bispectra, for a given configuration ($k_1 = 0.15h$~Mpc$^{-1}$, $k_2 = 0.25h$~Mpc$^{-1}$), for the emulated high-resolution (HR) simulations and corresponding reference high- and low-resolution (LR) realizations. The close agreement between the predictions and the ground truth implies that our super-resolution emulator reproduces with high fidelity the complex morphology, such as the filaments and voids, of the cosmic web.}
	\label{fig:bispectrum_plot}
\end{figure}

Gravitational evolution yields a non-Gaussian component to the cosmic matter structures on the small scales. These are key probes of the nature of gravity, dark matter and dark energy. To access the statistical information encoded in the non-Gaussian features, such as the peaks, filaments and voids, of the cosmic web, higher-order statistics (going beyond the power spectrum) are essential. We employ the bispectrum, the Fourier transform of the three-point correlation function, to quantify the spatial distribution of the cosmic structures, defined as:
\begin{equation}
    	(2\pi)^3 B(\myvec{k}_{1}, \myvec{k}_{2}, \myvec{k}_{3}) \delta_{\mathrm{D}}(\myvec{k}_{1}+\myvec{k}_{2}+\myvec{k}_{3}) = \langle \delta(\myvec{k}_{1}) \delta(\myvec{k}_{2}) \delta(\myvec{k}_{3}) \rangle , \label{eq:bispectrum}
\end{equation}    
where $\delta_{\mathrm{D}}$ is the Dirac delta.

\medskip
The bispectra reconstructed from the emulated high-resolution density fields and the reference simulations are depicted in Fig.~\ref{fig:bispectrum_plot}. We have chosen a particular configuration, $k_1 = 0.15h$~Mpc$^{-1}$, $k_2 = 0.25h$~Mpc$^{-1}$, and studied the variation with the angle between the two vectors. This choice, as justified in \citet{giusarma2019learning}, corresponds to the smallest scales typically used for cosmological parameter inference from measured galaxy power spectra. As can be seen by comparing the blue and red lines, the super-resolution emulator performs extremely well in reproducing the non-linear structures of the cosmic web.

\subsection{Void abundance}
\label{void_abundance}

\begin{figure}
	\centering
		{\includegraphics[width=\hsize,clip=true]{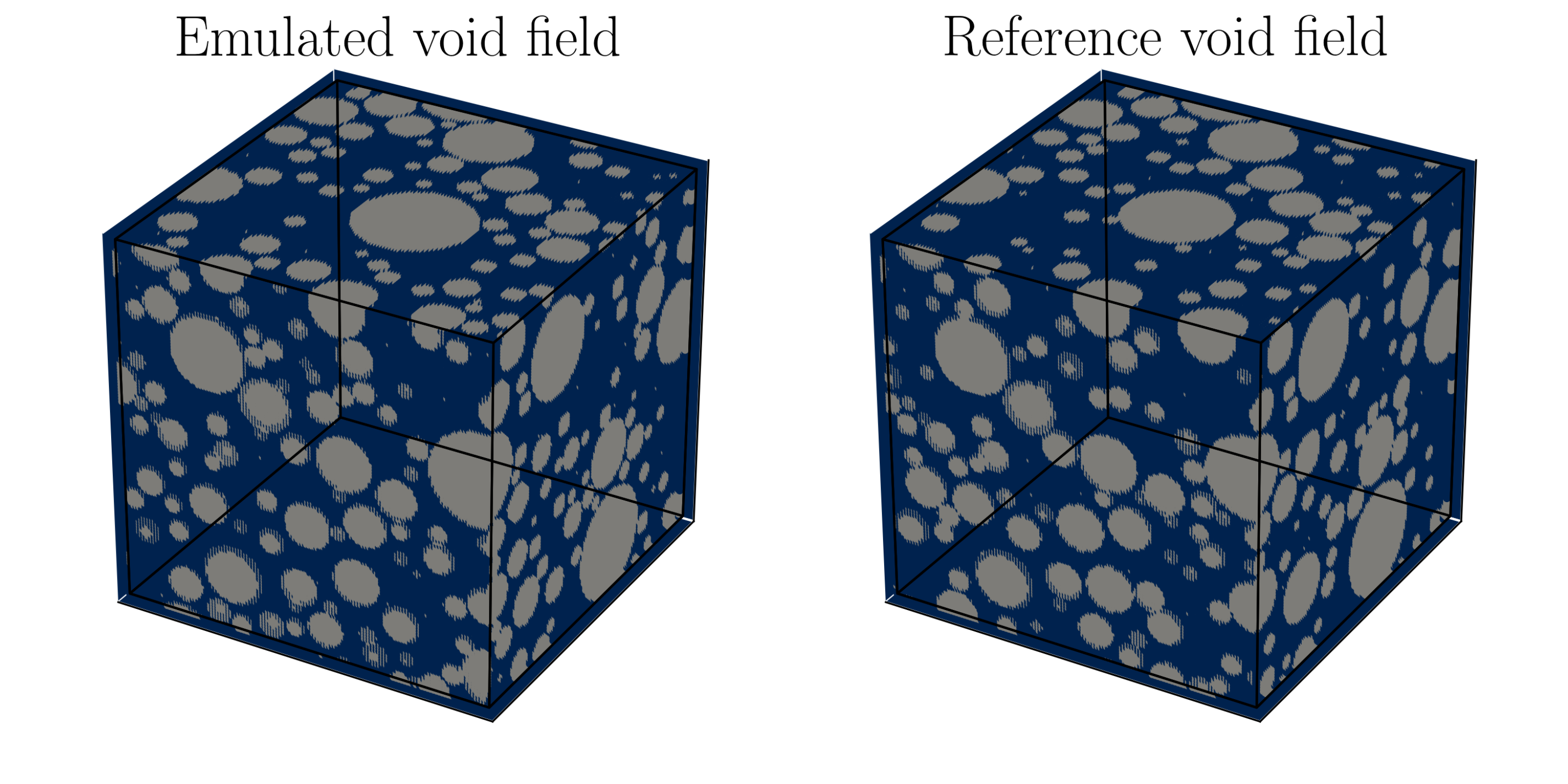}}
	\caption{3D void fields from the emulated high-resolution simulations and the corresponding reference density field for a 3D slice of length $500h^{-1}$~Mpc. Our super-resolution emulator adequately reproduces the properties of the void distribution. Whilst the general size and positions of the voids agree extremely well, the positions of smaller voids are occasionally slightly off.}
	\label{fig:void_field_3D}
\end{figure}

\begin{figure}
	\centering
		{\includegraphics[width=\hsize,clip=true]{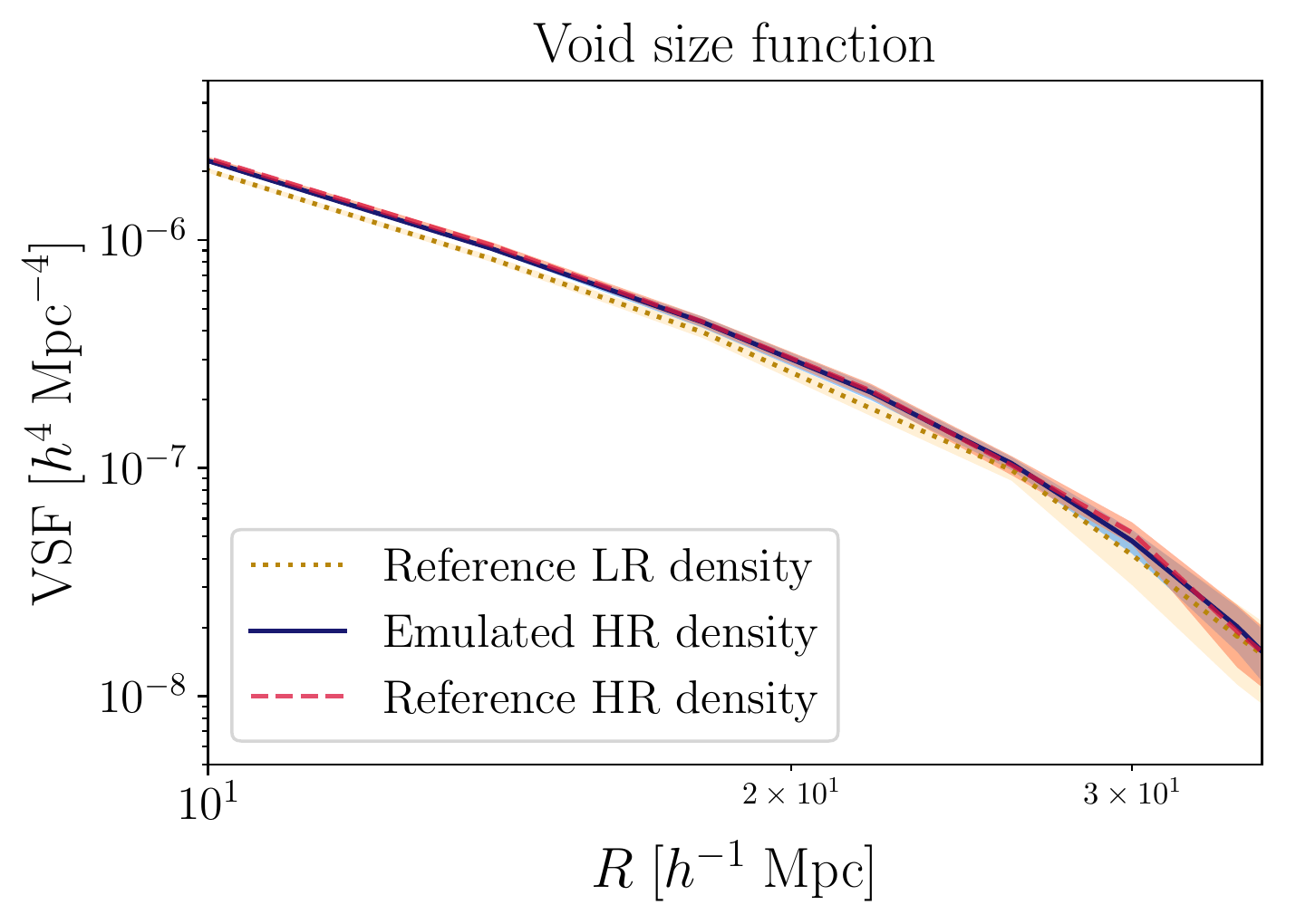}}
	\caption{Abundance of voids as a function of void radius, as described by the void size function, for the emulated super-resolution simulations and corresponding reference realizations. Our emulator accurately replicates the void abundance from the actual high-resolution (HR) simulations, and as such, fully captures the complex structures in the cosmic web.}
	\label{fig:VSF}
\end{figure}

We finally investigate the capacity of our super-resolution emulator to also replicate the void statistics in the actual high-resolution simulations. Voids are the under-dense regions of the cosmic web and encode a substantial amount of cosmological information \citep[e.g.][]{lavaux2012precision, hamaus2016constraints} by virtue of their ubiquity in the Universe. Two useful diagnostics, therefore, are the 3D void distribution and the void size function (VSF) describing the variation of the void abundance with void radius.

\medskip
To compute the VSF, we must identify the voids in the density fields. To that end, we adopt the void finding algorithm devised by \citet{banerjee2016simulating}, which involves smoothing the density field with a top-hat filter on a given scale $R$, initially chosen to be rather large, and selecting the voxels whose density is below a chosen threshold. The voxel, with the lowest density among the selected voxels, is then identified as the void centre with associated radius $R$. This procedure is repeated for subsequently smaller values of $R$, yielding a hierarchical identification of larger to smaller voids.

\medskip
Fig.~\ref{fig:void_field_3D} displays the reconstructed 3D void fields from the emulated high-resolution density field and the reference simulation in Fig.~\ref{fig:visual_inspection}. From a visual inspection, we find that the network adequately reproduces the 3D distribution of voids, although the positions of smaller-sized voids are slightly off in some places. This may be due to the model predictions on the smallest scales being limited by the grid resolution. The overall network performance is quantitatively verified by the respective VSF variations as shown in Fig.~\ref{fig:VSF}. This implies that the emulator accurately reproduces the void abundance in the actual high-resolution simulation.

\subsection{Dependence on cosmology}
\label{cosmology_dependence}

\begin{figure}
	\centering
		{\includegraphics[width=\hsize,clip=true]{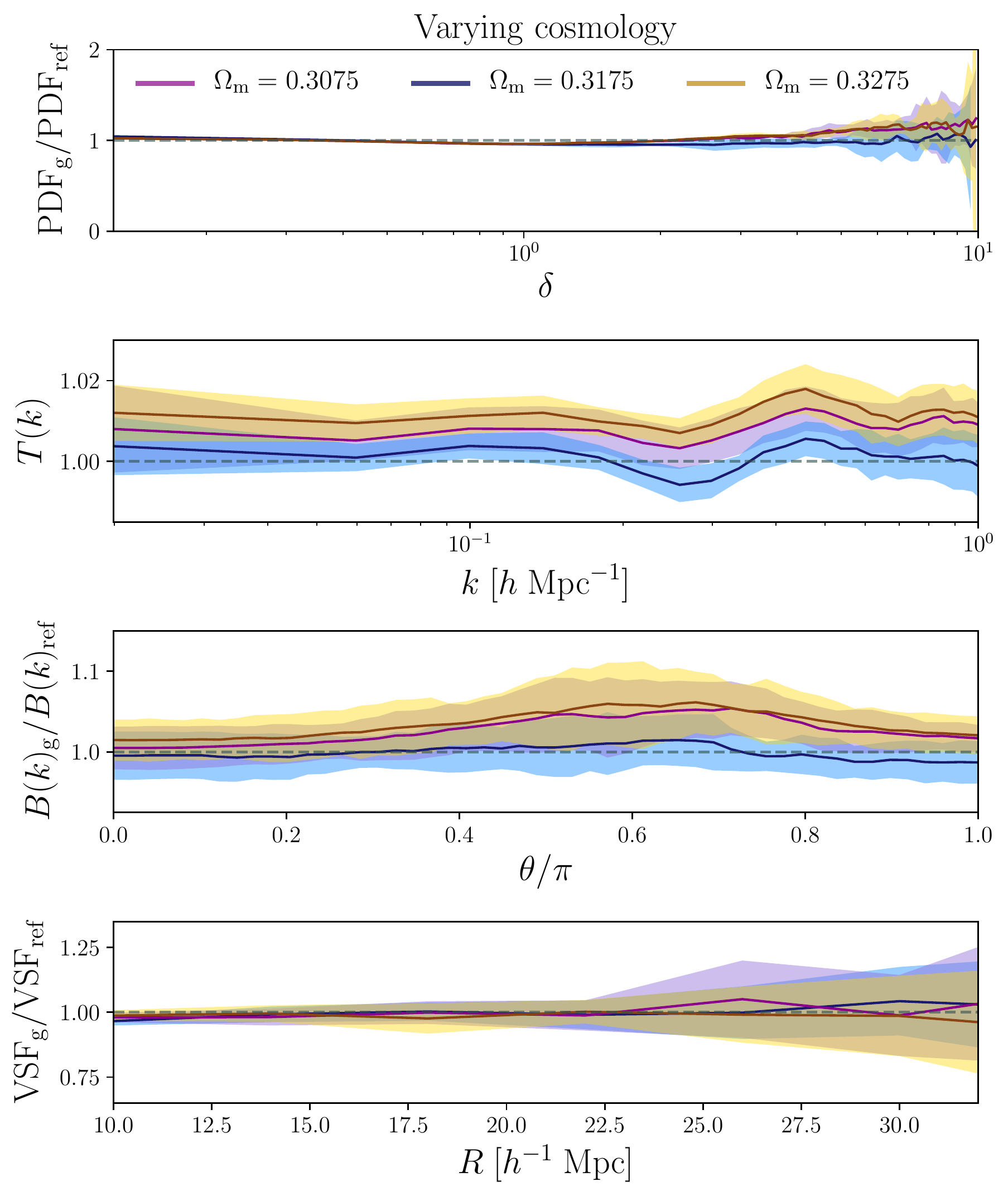}}
	\caption{Fractional deviations of the model predictions for the 1D PDF, power spectrum, bispectrum and VSF from their respective ground truth for emulated simulations at different cosmologies labelled in the top panel. The super-resolution emulator, optimized at a given fiducial cosmology, displays a slight bias at the level of the power spectrum and bispectrum when emulating simulations at varying cosmologies.}
	\label{fig:cosmo_dependence}
\end{figure}

We also verify the cosmological dependence of the super-resolution emulator, optimized at a given fiducial cosmology of $\Omega_\mathrm{m}=0.3175$, by applying it to two additional sets of ten low resolution simulations of cosmologies with $\Omega_\mathrm{m}=0.3075$ and $\Omega_\mathrm{m}=0.3275$. These two variants are obtained assuming different values of $\Omega_\mathrm{m}$, but with the initial random phases kept the same, such that only the physical effects induced by different cosmological parameters are introduced. We find that the super-resolution emulator is fairly robust to slight variations in cosmology, in contrast to the halo painting emulator from KCL19. Fig.~\ref{fig:cosmo_dependence} illustrates the various summary diagnostics with respect to their corresponding ground truth. The fractional deviations of the power spectrum and bispectrum are at the level of $2\%$ and $5\%$, respectively, with no substantial bias induced for the 1D PDF and VSF of the emulated simulations. This slight bias may result from the emulator applying an overall local transformation to the density field. One potential solution to mitigate this issue is to pass only the small-scale modes of the high-resolution initial conditions during the optimization procedure, such that the large-scale modes of the emulated field emanate directly from the low-resolution density field. This would render the emulator more robust to variations in cosmology, while also improving its overall performance. We defer this potential extension of our emulator to a future investigation.

\medskip
The execution time for the high-resolution simulation ($512^3$ voxels) is around 500 CPU hours, while the low-resolution one ($256^3$ voxels) required roughly 45 CPU hours. This implies a speed-up factor of $\sim 11$ when using the super-resolution emulator rather than running the high-resolution simulation. In terms of storage, using the emulator to generate the high-resolution simulation from the low-resolution one on the fly provides a storage gain by a factor of 8. This gain in computation time and storage would be even more significant when mapping to higher resolutions (for instance, $1024^3$ voxels), which is feasible using the super-resolution emulator. The network architecture may, however, require some tuning to optimize its performance for other resolution settings.

\subsection{Visualization of filters and feature maps}
\label{filter_visualization}

In an attempt to introspect the inner-workings of the super-resolution emulator and gain some physical understanding of the function the emulation is approximating, we visualize the filters of the convolutional kernels and their resulting feature maps. These indicate the informative features as inferred by the neural network and consequently provide some insights about the influence of the large-scale density distribution on the small-structures, although the conclusions that may be drawn from such visualizations are rather limited. We examine the filters and feature maps at the first layer of convolutions in the emulator where the majority of the information on the combination of different scales comes from.

\begin{figure*}
	\centering
    \subfloat[$7\times7\times7$ convolutional kernels and feature maps]{\includegraphics[width=\hsize]{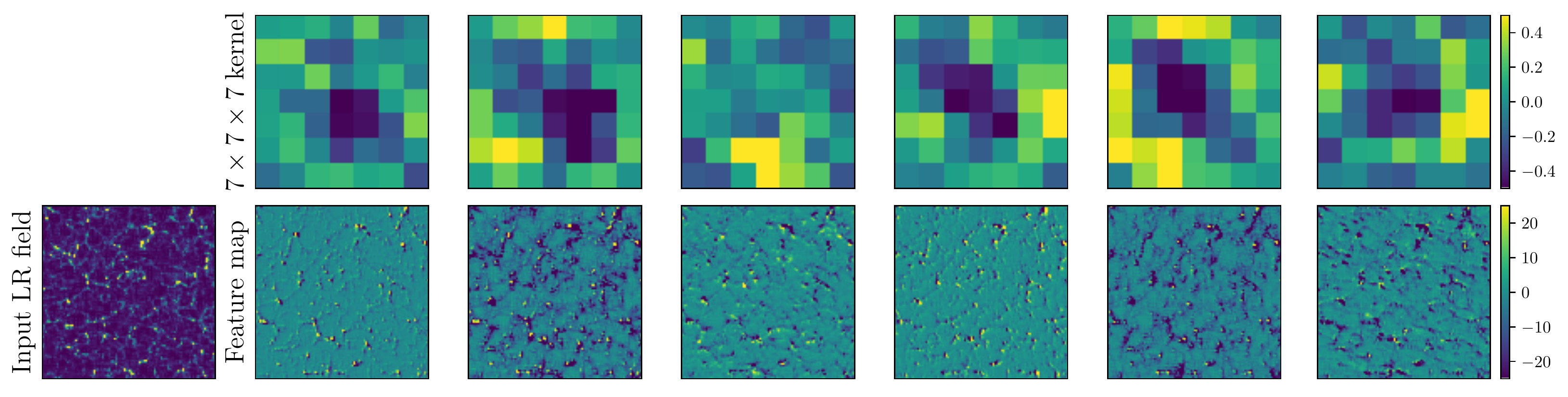}}
    \quad
    \subfloat[$5\times5\times5$ convolutional kernels and feature maps]{\includegraphics[width=\hsize]{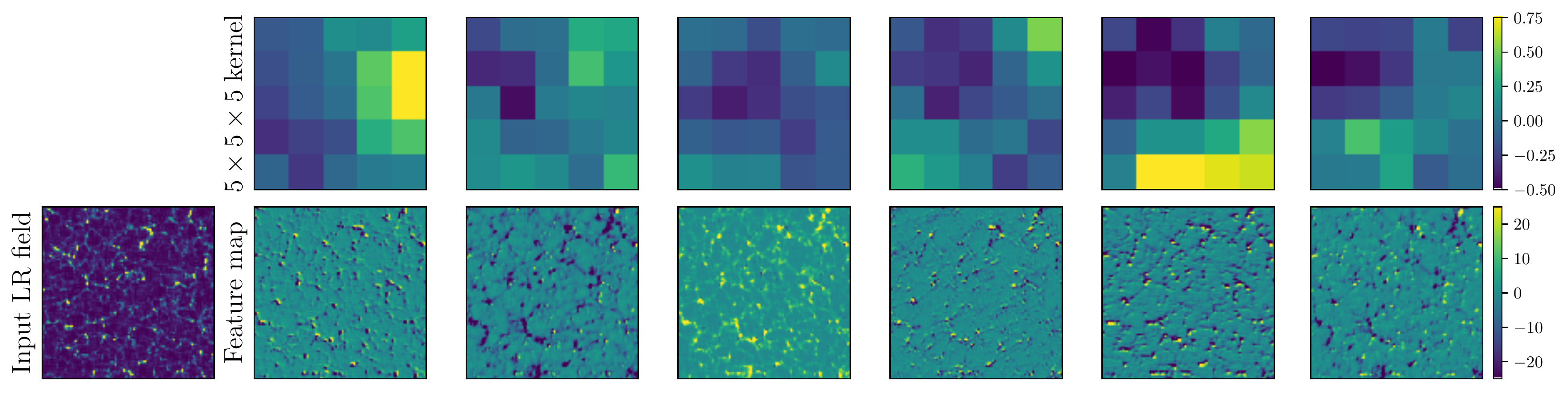}}
    \quad
    \subfloat[$3\times3\times3$ convolutional kernels and feature maps]{\includegraphics[width=\hsize]{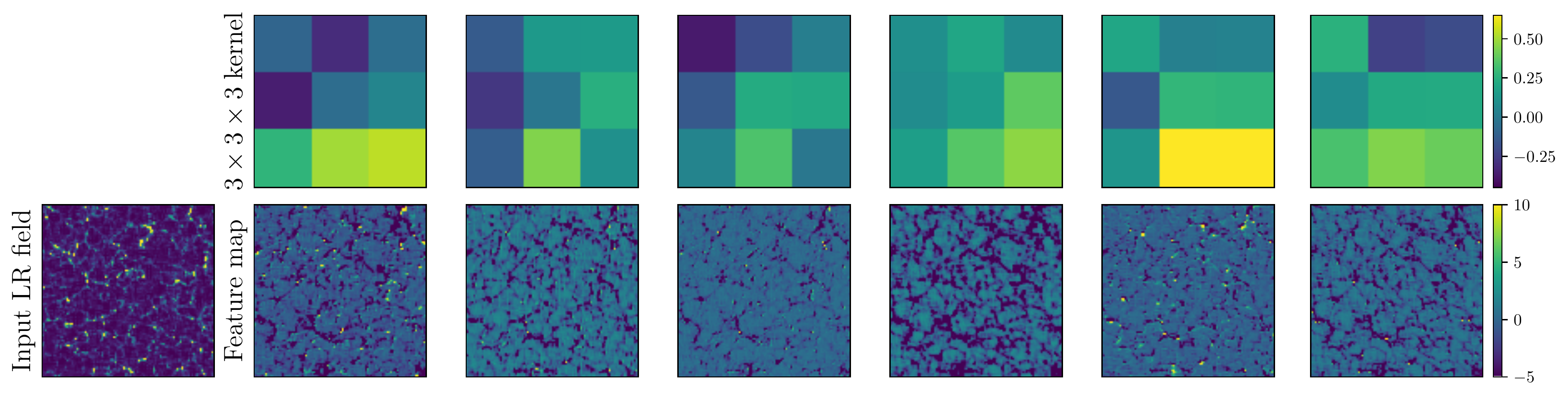}}
    \quad
	\caption{Convolutional kernels and corresponding feature maps in the top layer of the first residual Inception module for the low-resolution (LR) density field (cf. Fig.~\ref{fig:WGN_schematic}) for kernel sizes of (a) $7\times7\times7$, (b) $5\times5\times5$ and (c) $3\times3\times3$, from top to bottom, respectively. Note that the central slices of the 3D kernels and feature maps are illustrated, with the colour scale anchored for a given row. The feature maps in the bottom rows result from the convolution of the input low-resolution field shown with the kernels in the top rows. The kernels, in particular the $7\times7\times7$ ones, have an approximately radial distribution, implying that the network has learned a certain degree of rotational symmetry, as expected from the cosmological principle. The feature maps from the larger-sized kernels contain finer structures and in general encode the information about the peaks in the density field, whilst the smaller kernels try to enhance the contrast between the filaments and underdense regions.}
	\label{fig:kernels_feature_maps_LR}
\end{figure*}

\begin{figure*}
	\centering
    \subfloat[$7\times7\times7$ convolutional kernels and feature maps]{\includegraphics[width=\hsize]{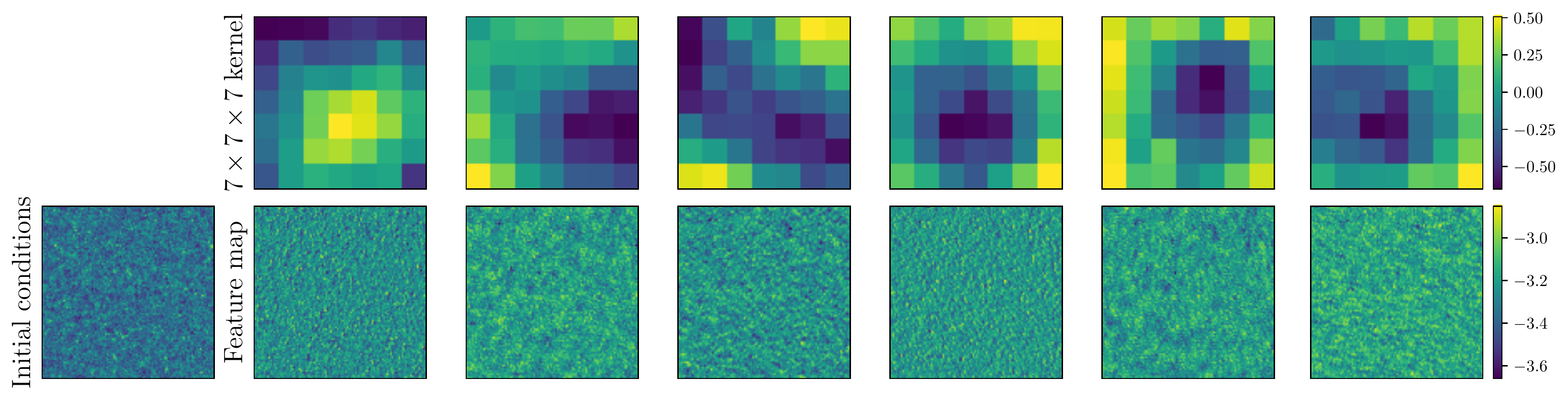}}
    \quad
    \subfloat[$5\times5\times5$ convolutional kernels and feature maps]{\includegraphics[width=\hsize]{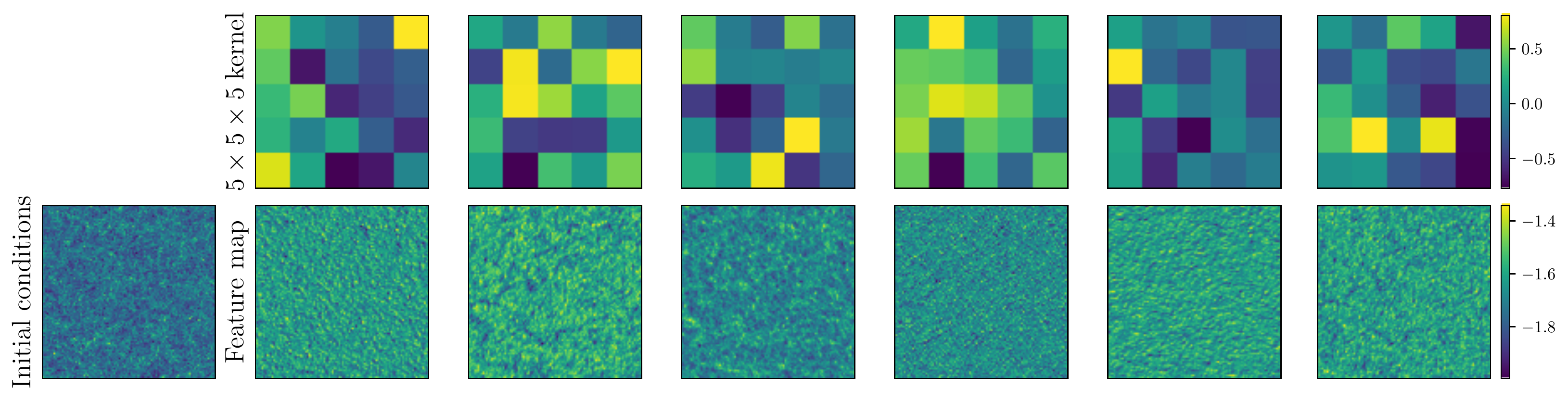}}
    \quad
    \subfloat[$3\times3\times3$ convolutional kernels and feature maps]{\includegraphics[width=\hsize]{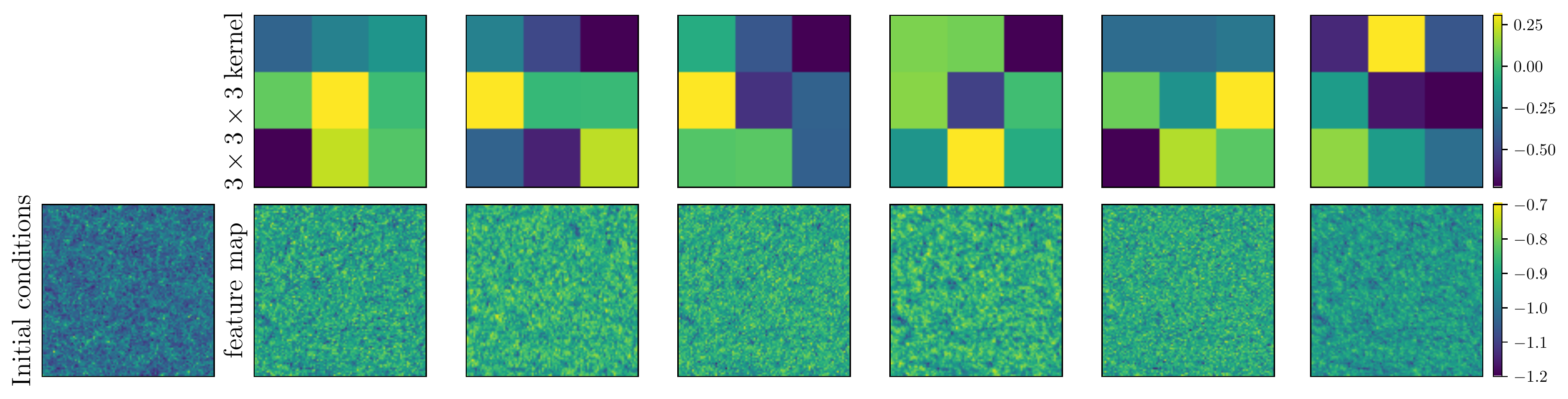}}
    \quad
	\caption{Convolutional kernels and corresponding feature maps in the top layer of the first residual Inception module for the input initial conditions for kernel sizes of (a) $7\times7\times7$, (b) $5\times5\times5$ and (c) $3\times3\times3$, from top to bottom, respectively. Note that the central slices of the 3D kernels and feature maps are illustrated, with the colour scale anchored for a given row. The initial conditions are also depicted for comparison. As for the low-resolution density field, the radial features are most informative for the larger kernels, but the weights are spread over a wider distribution, which characterizes the causal information encoded over the scales relevant for structure formation.}
	\label{fig:kernels_feature_maps_IC}
\end{figure*}

\medskip
Fig.~\ref{fig:kernels_feature_maps_LR} depicts the convolutional kernels and corresponding feature maps in the input layers of the residual Inception block for the low-resolution density field in the top and bottom rows, respectively, of each panel. The distinct kernel sizes in the Inception module are displayed in the different panels. Similarly, Fig.~\ref{fig:kernels_feature_maps_IC} illustrates the kernels and associated feature maps for the input initial conditions. The feature maps, in essence, yield the fields which the neural network has found most informative about the input initial conditions and low-resolution density distribution. As expected, we find that the larger kernels extract the large-scale features while the smaller sized kernels are pulling out the finer structures from the respective inputs. These feature maps are combined in non-linear fashion to provide the final emulated high-resolution density field. As a consequence of this non-linearity, the subsequent feature maps deeper into the network contain a high level of abstraction and are not illustrated here.

\medskip
The larger-sized $7\times7\times7$ kernels, for both the input low-resolution field and initial conditions, depicted in the top rows of Figs.~\ref{fig:kernels_feature_maps_LR} and \ref{fig:kernels_feature_maps_IC}, respectively, appear to have a radial distribution, thereby showing that the network has learned the rotational symmetry on the larger scales to a reasonable extent, as per the underlying fitting rationale outlined in Section~\ref{fitting_methodology}, driven by the cosmological principle. This, therefore, motivates the use of the multipole expansion of kernels, as employed in \citet{charnock2019neural} to use progressively stronger breaking of rotational symmetry as a principle for ordering kernel complexity, which would potentially further reduce the number of model parameters by orders of magnitude. In the case of the initial conditions, the weights are smeared over a broader distribution, highlighting the causal information encoded over the scales relevant for growth of cosmic structures. In general, the convolutional kernels try to enhance the contrast between the denser regions, such as the filaments, and the underdense ones, such as the voids.

\section{Conclusions and outlook}
\label{conclusions}

We have presented a super-resolution $N$-body emulator\footnote{The source code repository is available at \url{https://github.com/doogesh/super_resolution_emulator}.} using a deep physical modelling approach, with inspiration from our recent halo painting network \citep{DKR2019painting}. Our emulator, once optimized, allows us to populate dark matter simulations with high-resolution structures in a fraction of a second on a modern GPU. We showcased the performance of our method in reproducing the intricate filamentary pattern of the cosmic web from a ground truth simulation using several reliable diagnostics. It should be clearly noted that this emulator is designed to provide high quality, deterministic approximations of the true high-resolution simulations that are not necessarily realizations of the true distribution of the simulations as could, in the ideal case, be emulated using generative techniques. We note that we have not attempted such work here since there is no known current methods that guarantee that generated realizations are really samples from the true distribution of data, and we preferred high quality, interpretable emulation over arbitrary generation of samples from black boxes.

\medskip
The network design and architecture are driven by physical principles, in contrast to a conventional black box approach. We have focused on translational and causal information with a good approximation to rotational symmetry to finesse our network. As such, our physically motivated emulator has relatively few network parameters and may be fit sufficiently well using a single set of simulations. Recent deep generative models, as proposed in the cosmology community, rely on roughly three orders of magnitude more network parameters, which require several thousands of simulations for training. Moreover, since convolutional kernels are translationally invariant, our neural network can emulate high-resolution simulations of arbitrary box size, for a given physical voxel size, allowing us to cheaply generate extremely large, high-resolution simulations. The large-scale modes which are not present explicitly in the data used for the fit will still be present in the super-resolution large size box since they are provided by the low-resolution simulation. We have yet to implement true rotational symmetry which has now been considered in~\cite{charnock2019neural}, which could potentially lead to orders of magnitude fewer model parameters for the fit.

\medskip
Such an emulator of cosmic dynamics may be employed as a means to accelerate the statistical inference framework of algorithms based on Bayesian forward modelling approaches \citep[e.g.][]{jasche2013bayesian, jasche2019physical, DKR2018altair, porqueres2019inferring}, to render high-resolution analyses of upcoming galaxy surveys feasible. In this context, combining neural physical engines with forward modelling techniques is an extremely promising avenue for next-generation cosmological analyses \citep{charnock2019neural}. Another practical use of such an emulator is to create accurate approximate realizations of high-resolution simulations on the fly using stored low-resolution simulations and generating only the high-resolution initial conditions. For the given low and high resolutions considered in this work, the gain in execution time and storage capacity is by roughly an order of magnitude, which is significant for applications requiring several thousands of simulations.

\medskip
Upcoming galaxy surveys, such as LSST and Euclid, will cover unprecedented volumes of the sky. While it is computationally cheap to run $N$-body simulations extending over such cosmological volumes, it is not feasible to run such simulations at a sufficiently high-resolution to resolve the smallest halos containing the galaxies expected to be observed by these surveys. Our super-resolution emulator, therefore, constitutes the first step in addressing this computational bottleneck using the latest advances in generative modelling techniques. Achieving this ultimate goal would naturally require an emulator capable of mapping to higher resolutions than considered in this work. This would also lead to other practical applications involving the estimation of covariance matrices and light cone production from mock observations of the Universe. The emulator may also be used to generate fast high-resolution simulations to be fed to the recently proposed information maximizing neural network \citep[IMNN,][]{charnock2018automatic} to learn the optimal function that summarizes the data, with the trained IMNN subsequently being utilized in a likelihood-free inference setting. By feeding the IMNN with high-resolution simulations emulated from the same low-resolution field, but different initial conditions, the network will learn to ignore the small-scale fluctuations and produce summaries that are robust to small-scale non-linearities \citep[analogous to the nuisance-hardened summaries in][]{alsing2019nuisance}.

\medskip
Finally, we have so far only considered using the emulator to give us approximations to the data via a simple physically motivated fitting procedure. It would be interesting to incorporate a probabilistic model of how likely features are to exist in the data which would allow us to really sample from the known distribution of data. In this procedure, one would fit a conditional pixelwise probability estimator \citep{lanusse2019pixel}, but rather than using the output as a probabilistic function, the uncertainties in the fitting could be pushed back to the feature kernels. By analyzing how likely any kernel is, we will be able to understand physically how the interactions between the low-resolution density field and high-resolution initial conditions provide us with the structures that appear in the high-resolution simulations. Since the number of weights in this model is still relatively large, we would propose using variational inference-type fitting, which is becoming very promising using frameworks such as \textsc{TensorFlow Probability} \citep{dillon2017TFP}. Note that here we are looking for indications of uncertainty in the kernels to learn about the physical processes, rather than trusting the probabilistic interpretation of the network for true inference, for which we would require more direct sampling schemes, such as those used in \citet{charnock2019neural}, and a good comprehension of the likelihood of the weights in the model. However, once fitted, realizations of the approximate high-resolution images could be obtained for many samples from the uncertain kernels, and the mean and variance high-resolution maps could be produced, describing where best to trust and not to trust the approximate function.

\section*{Acknowledgements}

We thank the reviewer for their constructive feedback which helped to improve the quality of the manuscript. We express our appreciation to Guilhem Lavaux for his valuable insights. We acknowledge financial support from the ILP LABEX, under reference ANR-10-LABX-63, which is financed by French state funds managed by the ANR within the Investissements d'Avenir programme under reference ANR-11-IDEX-0004-02. DKR is a DARK fellow supported by a Semper Ardens grant from the Carlsberg Foundation (reference CF15-0384). Part of the work of FVN and BDW has been supported by the Simons Foundation. TC is supported by the ANR BIG4 project, grant ANR-16-CE23-0002 of the French Agence Nationale de la Recherche. TC would like to thank NVIDIA for the Quadro P6000 used in this research. The bispectrum and void finding computations were performed using the publicly available \textsc{pylians}\footnote{\url{https://github.com/franciscovillaescusa/Pylians}} code. This work is done within the Aquila Consortium.\footnote{\url{https://aquila-consortium.org}}




\bibliographystyle{mnras} 
\bibliography{./compiled_references} 




\bsp	
\label{lastpage}
\end{document}